\documentclass[9pt,twocolumn,twoside]{osajnl}
\usepackage{balance}
\journal{josaa} 

\setboolean{shortarticle}{true}

\title{Circular Airy vortex beams carrying two point vortices symmetrically on x-axis with same or opposite signs}

\author[1]{Zhifang Qiu}
\author[2,4]{Bingsong Cao}
\author[1,5]{KaiKai Huang}
\author[3]{Xian Zhang}
\author[1,3,6]{Xuanhui Lu}

\affil[1]{Physics Department, Zhejiang University, Hangzhou 310027, China}
\affil[2]{Department of Physics, Huzhou University, Huzhou 313000, China}
\affil[3]{Advanced Technology Institutes, Zhejiang University, Hangzhou 310027, China}

\affil[4]{e-mail: bingsongcao@zju.edu.cn}
\affil[5]{e-mail: huangkaikai@zju.edu.cn}
\affil[6]{e-mail: xhlu@zju.edu.cn}

\begin{abstract}
	We study circular Airy vortex beams (CAVBs) with two same or opposite point vortices symmetrically loaded on $x$-axis, inside, onside and outside of the main ring (The absolute value of topological charge is one). It is found that the loading position (denoted as $d_v$) can significantly tune the auto-focusing behaviour. For two vortices with same signs, when $d_v$ is near zero, the double vortices approximately merge at the focal plane. But as $d_v$ increases, the merged vortices will gradually separate along the y-axis. When $d_v$ is large, the focal spot becomes Bessel-shaped pattern. For opposite vortices, when $d_v$ is small, the focusing pattern is Bessel-shaped (no vortex exists in the centre) but not strictly in the centre. With the increase of $d_v$, the original spot gradually moves up along the $y$-axis and transfers its energy to the newborn spot below $x$-axis, at the same time, the newborn spot moves towards the centre. When $d_v$ is large, the focus returns to a Bessel-like pattern. Good agreements have been achieved between simulations and experiments. The tailoring effect of $d_v$ of the two vortices on the focal pattern may have potential applications in micromanipulation, communication and other fields.
\end{abstract}

\setboolean{displaycopyright}{true}

\begin{document}

\maketitle

\section{Introduction}

Since first proposed in 1979 by M. V. Berry and N.L. Balazs \cite{doi:10.1119/1.11855}, Airy beams\cite{Siviloglou:07,recent_advances_Efremidis:19,Zhuang:15,Chu.2012} have been widely studied due to their unique properties of non-diffractive, self-healing, and self-bending. Through radial symmetry, circular Airy beam (CAB) or called the abruptly autofocusing (AAF) or Ring Airy beam (RAB) was first introduced in 2010 \cite{Efremidis:10} , extended through pre-engineering in 2011 \cite{Pre-engineered-Chremmos:11}, and first generated experimentally in 2011 \cite{Papazoglou:11}. Without the use of lenses or non-linearity, CAB will abruptly autofocus in the focus area while maintaining a low intensity until that position. This feature is very useful in the field of particle capture\cite{Chremmos:11,Zhang:s,Jiang:s1,Jiang:s2,Wang2014}, optical bullet generation\cite{Panagiotopoulos2013}, laser biomedical treatment or material process\cite{Manousidaki:s}, atomic manipulation and other fields\cite{Papazoglou:11}.

Optical vortices, with unique spiral phase distribution, carrying optical angular momentum (OAM), existing null intensity around phase singularity, have attracted wide attention since 1989\cite{COULLET1989403} and aroused many applications, such as optical manipulation\cite{He.1995,Gahagan.1996,ZhongxiWang.2011}, communication\cite{Gibson.2004,Lavery.2017}, high resolution imaging and measurement\cite{G.Anzolin.2009}. The airy vortex beam which combines the advantages of Airy beam and vortex beam is an interesting research object with a lot of potential applications and in recent years it has received many attentions. Airy beam carrying optical vortex was first proposed by Mazilu\cite{MichaelMazilu.2009} in 2009, then studied and generated by Dai \cite{Dai:10,Dai:11}, and studied by Chen\cite{Chen.2011,Chen.2012}.  In 2012, autofocusing beams with vortices were generated by Davis \cite{Davis:12,Davis:13-3}. 

Here come to the problem of circular Airy vortex beam carrying some optical vortices (OVs), it was first studied by Jiang \cite{Jiang:12} in 2012. In his article\cite{Jiang:12} (the OV is r-type), Jiang studied the CAVB with on-axis OV, off-axis OV and two symmetrically loaded OVs. The OVs in these cases are all loaded inside the main ring of the background CAB. In his study of two OVs, topological charge $m$ being same and opposite are all included ($\left| m \right|=1$). Some similar problems have been studied in recent years. In 2015, sharply autofocused ring Airy Gaussian vortex beam\cite{Dongmei-Chen:s} was proposed and the authors studied a pair of positive r-type OVs loaded inside main ring. In 2018, Jiang\cite{Jiang:s33} studied the CAVB with two opposite r-type OVs under different initial launch angles and the OVs are loaded inside main ring. In 2020, Zhuang\cite{JingliZhuang.2020} studied tight-focusing properties of linearly polarized circular Airy Gaussian vortex beam loaded with two point-type OVs with same topological charges (equals  1), and the position of the two OVs are inside main ring. Recently, Zhang et al. \cite{XiangZhang.2020} pointed out that the loading position of the OV has an important influence on the autofocusing behaviour of the CAVB. The authors studied the effects of loading position of the off-axis vortex (near the main ring of the CAB), the vortex type and the topological charge on the autofocusing behaviour. From the above contents, we are sure that no one has done the exploration of CAVBs carrying two vortices with same and opposite topological charges, symmetrically located inside, onside and outside of the main ring. In fact, the idea of exploring the CAVBs with this case has been around us long before we saw this article by ZHANG\cite{XiangZhang.2020}. We here study the effect of the loading position of a pair of OVs on the characteristics of the intensity pattern as well as the vortices distribution at the autofocusing plane, and the results are quite interesting.

This manuscript is organized as follows. In Section 2, we numerically study the characteristics of the beam at autofocusing plane changing with the loading position $d_v$ of two vortices. In Section 3, we experientially study this relationship with $d_v=0.5,1,1.5 R_0$ through a phase-only SLM with a 4f system\cite{Mendoza-Yero:14} ($R_0$ denotes the main ring radius of the background CAB at initial plane). In Section 4, we summarize the unique tailoring effect of the parameter $d_v$ on the focal pattern of the CAVBs loaded with two vortices symmetrically and briefly discuss some potential applications.

\section{Simulating results}

The electric field of the linear polarized CAVBs at the initial plane with a pair of point-type vortices can be defined as\cite{Jiang:12,D.R.1997}:

\begin{equation}
\label{eq1}
\!U\!\left( {\!x,\!y,\!z \!= 0} \right) \!= \!C \!\times {\!\text{Ai}}\!\left( {\!\frac{{{r_0} \!- \!r}}{w}} \!\right)\!\exp \!\left( {\!a\!\frac{{{\!r_0} \!- \!r}}{w}} \!\right)\!\exp \!\left( {i{m_1}{\varphi _1}} + {i{m_2}{\varphi _2}} \!\right), 
\end{equation}
where Ai is the Airy function; $r_0$ is related to the radial position of the main Airy ring at the initial plane; $w$ is the radial scale; $a$ is the decaying parameter; $C$ is a constant related to beam power; $m_1,m_2$ are the topological charges of the two vortices, where $\varphi_1$ and $\varphi_2$ are calculated from:
\begin{equation}
\label{eq2}
\begin{gathered}
{\varphi _1} = {\text{angle}}\left( {x - {x_1} + i\left( {y - {y_1}} \right)} \right) \hfill \\
{\varphi _2} = {\text{angle}}\left( {x - {x_2} + i\left( {y - {y_2}} \right)} \right) \hfill \\ 
\end{gathered},
\end{equation}
which means that the centre of two vortices with charges $m_1$ and $m_2$ are located at $(x_1,y_1)$ and $(x_2,y_2)$ respectively. The function "angle"  represents the phase angle of the complex number. According to reference\cite{Panagiotopoulos2013}, the radius of the main ring can be calculated as:
\begin{equation}
\label{eq3}
{R_0} = {r_0} - w*g\left( a \right),
\end{equation}
where $g(a)$ denotes the first zero point of the following equation in the region of $x<0$:
\begin{equation}
\label{eq4}
\frac{{d\operatorname{Ai} \left( x \right)}}{{dx}} + a \times \operatorname{Ai} \left( x \right) = 0. 
\end{equation}
The complex amplitude of the beam at any plane after the initial plane can be calculated by the angular spectrum formulas as follows:

\begin{equation}
\label{eq5}
\!U\!(\!x,\!y,\!z\!) \!= \!\frac{1}{{{{\left( {2\pi } \right)}^2}}}\!\iint {\!A\!\left( {{k_x},{k_y}} \!\right)}\!\exp \!\left( {i{k_z}z} \!\right) \!\times {\!\text{exp}}\!\left[ {\!i\!\left( {{k_x}x + {k_y}y} \!\right)} \!\right]\!d{k_y}d{k_x},
\end{equation}
where $k_0=2\pi/\lambda_0$ is the wave number, $\lambda_0$ is the vacuum wavelength, $k_z=\sqrt {{k_0}^2 - {k_x}^2 - k_y^2}$ is the z component of the wave vector $\overset{\lower0.5em\hbox{$\smash{\scriptscriptstyle\rightharpoonup}$}} {k}$ of the plane wave component, and the complex amplitude of the plane wave component $A\left({{k_x},{k_y}} \right)$ is determined by:
\begin{equation}
\label{eq6}
A\left( {{k_x},{k_y}} \right) = \iint {U\left( {x,y,0} \right)}\exp\left( { - i\left( {{k_x}x + {k_y}y} \right)} \right)dxdy.
\end{equation}
The abruptly autofocusing position $z_f$ for CAVB can be approximated by \cite{Li:14,XIE2018288}:
\begin{equation}
\label{eq7}
{z_f} \approx 2{k_0}{w^2}\sqrt {\frac{{{r_0}}}{w} + 1.018}.
\end{equation}

In the following part of this article, we choose $\lambda_0=1064$ nm, $r_0=1$ mm, $a=0.1$, $w=0.15$ mm, then according to Eq. (\ref{eq3}), (\ref{eq4}) and (\ref{eq7}) we have $R_0=1.14$ mm and $z_f=0.732$ m. The two vortices are located symmetrically on the $x$-axis with $x_1=-d_v$ and $x_2=d_v$. In the following simulation, we set the beam power equal to 1 W. For a pair of same vortices, we set $m_1=m_2=-1$, and for a pair of opposite ones, we set $m_1=-1$, $m_2=1$, and the value of $d_v$ is among $[0.5R_0,R_0,1.5R_0,2R_0,4R_0]$.

\begin{figure}[htbp]    
	\centering\includegraphics[width=\linewidth]{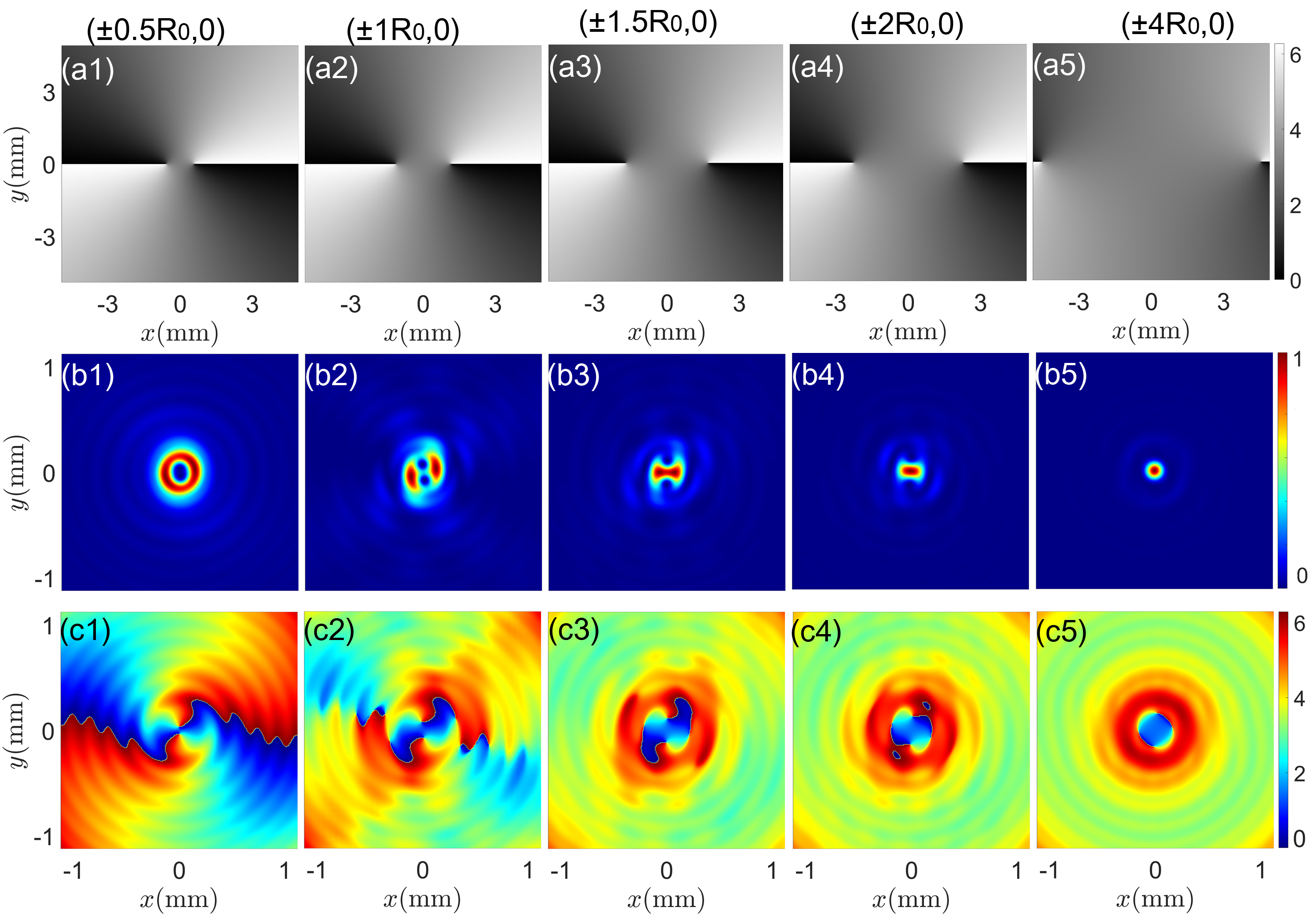}
	\caption{CAVBs carrying symmetrically two same vortices ($m_1$ = $m_2$ = -1) on $x$-axis at initial plane. (a) The phase generated from two vortices at initial plane. (b) the intensity pattern of the $z=z_f$ plane; (c) the phase pattern of the $z=z_f$ plane. The positions of two vortices of each row is labelled on the left side. (color online)}
	\label{fig-1}
\end{figure}

\begin{table}[htbp]
	\centering
	\caption{\bf locations of the old and newborn vortices in the focal plane changing with $d_v$, for CAVBs carrying a pair of same vortices located symmetrically on $x$-axis of the initial plane. (measured in simulation)}
	\begin{tabular}{cccccc}
		\hline
		&  $d_v$   &         evolved vortices         &     newborn vortices      &  &  \\ \hline
		& 0.5$R_0$ &   $( 0,\pm 32\text{um} )$    &             -             &  &  \\ \hline
		& 1.0$R_0$ &   $( 0,\pm 72\text{um} )$    &             -             &  &  \\ \hline
		& 1.5$R_0$ & $( {0,\pm 104{\text{um}}} )$ & $(\mp216\text{um},\pm40\text{um} )$    &  &  \\ \hline
		& 2.0$R_0$ & $( {0,\pm 136{\text{um}}} )$ & $(\mp184\text{um},\pm72\text{um})$ &  &  \\ \hline
		& 4.0$R_0$ & $( {0,\pm 152{\text{um}}} )$ & $(\pm152\text{um},\pm16\text{um} )$ &  &  \\ \hline
	\end{tabular}
	\label{table_same}
\end{table}

Fig.\ref{fig-1} shows the focusing properties of CAVBs carrying two point vortices with $m_1$ = $m_2$ = -1 on the $x$-axis of the initial plane. From first column to fifth column, the $d_v$ increases and the positions of the vortices are labelled in each column on the top of the figure. Row (a) denotes the phase generated by the two vortices, rows (b) and (c) represent the intensity pattern and the total phase distribution of the beam at $z=z_f$ plane respectively. From Fig.\ref{fig-1} (b1, c1), where $d_v=0.5R_0$, it can be seen that two vortices with same topological charges are forced to the beam centre at the focal plane, forming almost one vortex of charge -2. The two vortices at the focal plane are not completely overlapping, and we measured out in simulation that they are approximately located at (0,32um) and (0,-32um) respectively. The second column represents the case when a pair of point vortices locates on the main ring, that is, $d_v = R_0$. Compared Fig.\ref{fig-1} (c2) with (c1), one can see that the position of the two vortices formed at the focal plane is farther from the origin of the coordinates, and the locations measured in simulation are approximately (0,72um) and (0,-72um) respectively. Comparing Fig.\ref{fig-1} (b2) with (b1), it can be clearly seen that the intensity distribution at the focal plane changes accordingly, and there are two separate dark spots in the centre of Fig.\ref{fig-1} (b2), whose positions are probably consistent with that of the two vortices of Fig.\ref{fig-1} (c2). Similarly, from third column to fifth column, it can be seen that as $d_v$ increases, the two vortices (charges equal to -1) forming at the focal plane gradually moves away from the centre with creation of two new vortices (charges equal to 1), and it can be inferred from Fig.\ref{fig-1}(c5) that these four vortices finally would distribute onto a circle with limited radius and annihilate together, and the light intensity distribution finally returns to a circle (Bessel-shaped) pattern without dark region. In the process of increasing $d_v$, the approximate positions (measured in simulation) of the old and newborn vortices on the autofocusing plane are listed in Table \ref{table_same}. Where come the new vortices and how disappear the four vortices are worthy of further research.

\begin{figure}[htbp]  
	\centering\includegraphics[width=\linewidth]{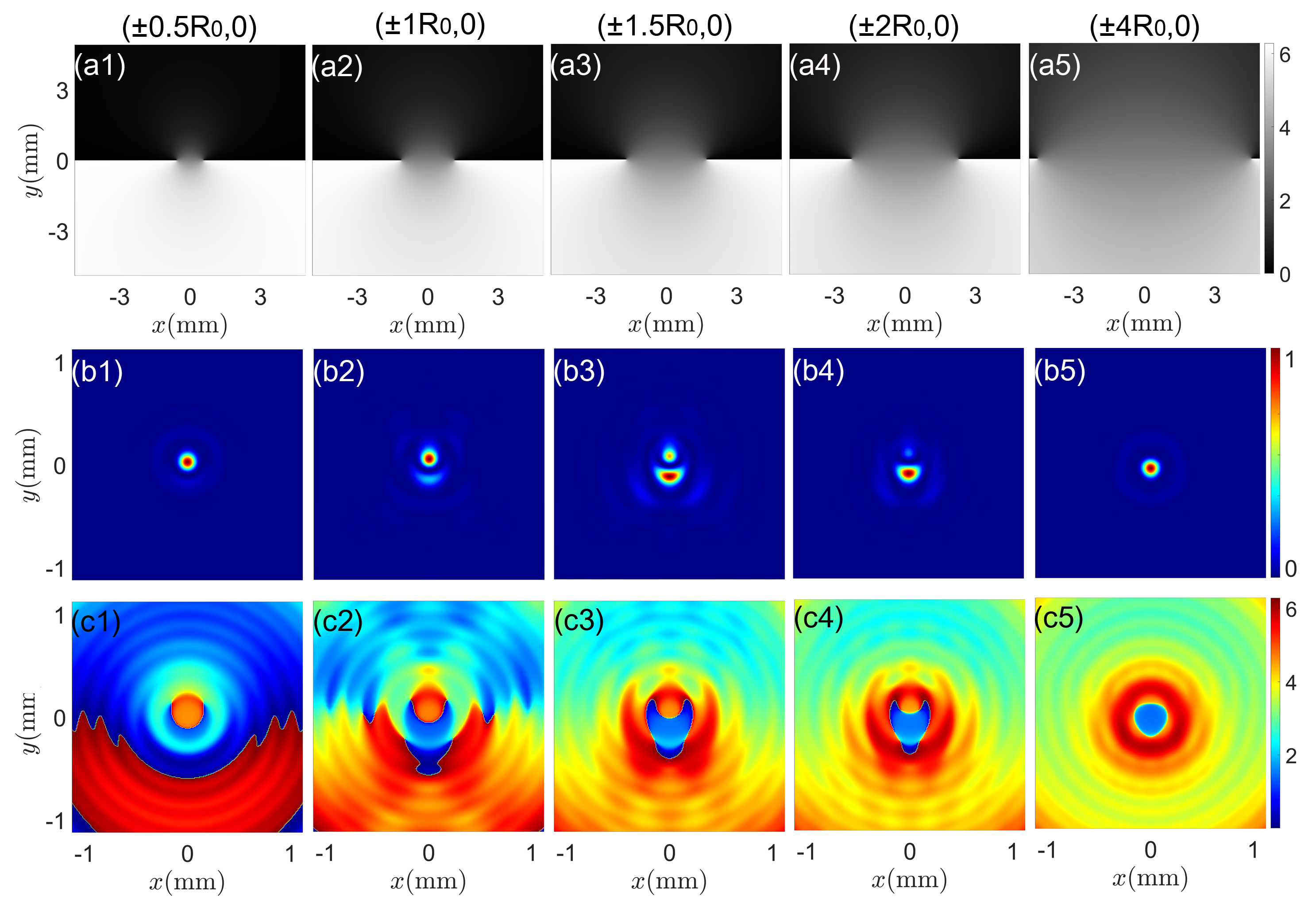}  
	\caption{CAVBs carrying symmetrically two opposite vortices on the $x$-axis at initial plane, the OV at ($-d_v$,0) has $m_1=-1$ and the other one has $m_2$=1. (a) The phase generated from two vortices at initial plane. (b) the intensity pattern of the $z=z_f$ plane; (c) the phase pattern of the $z=z_f$ plane. The positions of two vortices of each row is labelled on the left side. (color online)}
	\label{fig-2}
\end{figure}

Fig.\ref{fig-2} shows the focusing properties of CAVBs carrying two point vortices with $m_1$ = -1, $m_2$ = 1 on the $x$-axis of the initial plane. From first column to fifth column, the $d_v$ increases and the positions of the vortices are marked in each column on top of the figure. Row (a) denotes the phase generated by the two vortices, rows (b) and (c) represent the intensity pattern and the total phase distribution of the beam at $z=z_f$ plane respectively. From Fig.\ref{fig-2} (b1, c1), where $d_v=0.5R_0$, it can be seen that in the focal plane, the two vortices with opposite charges almost overlap and annihilate\cite{Jiang:12} totally at the beam centre, where there is a solid circular (Bessel-shaped) spot. Actually, the solid spot is located at about $(0,27\text{um})$, not at the origin of the coordinates and in Fig.\ref{fig-2} (c1), there do exist a pair of opposite vortices with order $\pm$1 at about $( \pm 160\text{um}, 136\text{um})$. In Fig.\ref{fig-2} (c2-c4), there exist similar vortices which locate symmetrically about the y-axis with opposite signs.
From (b1,b2,b3) of Fig.\ref{fig-2}, it can be seen that as $d_v$ increases, the bright spot at the origin of coordinates gradually shifts upwards along $y$-axis, accompanied by a bright spot gradually appears below the $x$-axis, and the energy is also gradually transferred to the newborn spot. At the same time, the region near the origin becomes dark, as shown in Fig.\ref{fig-2} (b3). As shown in Fig.\ref{fig-2} (b4), when $d_v$ increases to 2$R_0$, the energy has been mainly distributed in the bright spot below the $x$-axis, which tends to move towards the $x$-axis; As shown in Fig.\ref{fig-2} (b5), when $d_v$ increases to 4$R_0$, the autofocusing pattern returns to Bessel-shaped spot. However the solid spot is still not completely coincident with the origin and located at about $(0,-54\text{um})$. Table \ref{table_opposite} lists the locations of the brightest spot above and below the $x$-axis in the focal plane changing with $d_v$.

\begin{table}[t]
	\centering
	\caption{\bf locations of the brightest point of the spot above and below the $x$-axis in the focal plane changing with $d_v$, for CAVBs carrying a pair of opposite vortices located symmetrically on $x$-axis of the initial plane. (measured in simulation)}
	\begin{tabular}{cccccc}
		\hline
		 &  $d_v$   &   spot above $x$-axis   &     spot below $x$-axis     &  &  \\ \hline
		 & 0.5$R_0$ &   $( 0,27\text{um} )$   &       barely visible        &  &  \\ \hline
		 & 1.0$R_0$ &   $( 0,53\text{um} )$   & $( {0, - 146{\text{um}}} )$ &  &  \\ \hline
		 & 1.5$R_0$ & $( {0,77{\text{um}}} )$ & $( {0, -122{\text{um}}} )$  &  &  \\ \hline
		 & 2.0$R_0$ & $( {0,96{\text{um}}} )$ &  $( {0,-101{\text{um}}} )$  &  &  \\ \hline
		 & 4.0$R_0$ &     barely visible      &  $( {0, -54{\text{um}}} )$  &  &  \\ \hline
	\end{tabular}
	\label{table_opposite}
\end{table}

Interestingly, from Fig.\ref{fig-1} or Fig.\ref{fig-2}, one can easily find that the symmetry of the beams in both cases remain unchanged during their propagation process, all the graphs in Fig.\ref{fig-1} have a point reflection symmetry about the origin, and all graphs in Fig.\ref{fig-2} are axisymmetric about the y axis. 

\section{Experimental results}

\begin{figure}[h]	
	\centering\includegraphics[width=6.5cm]{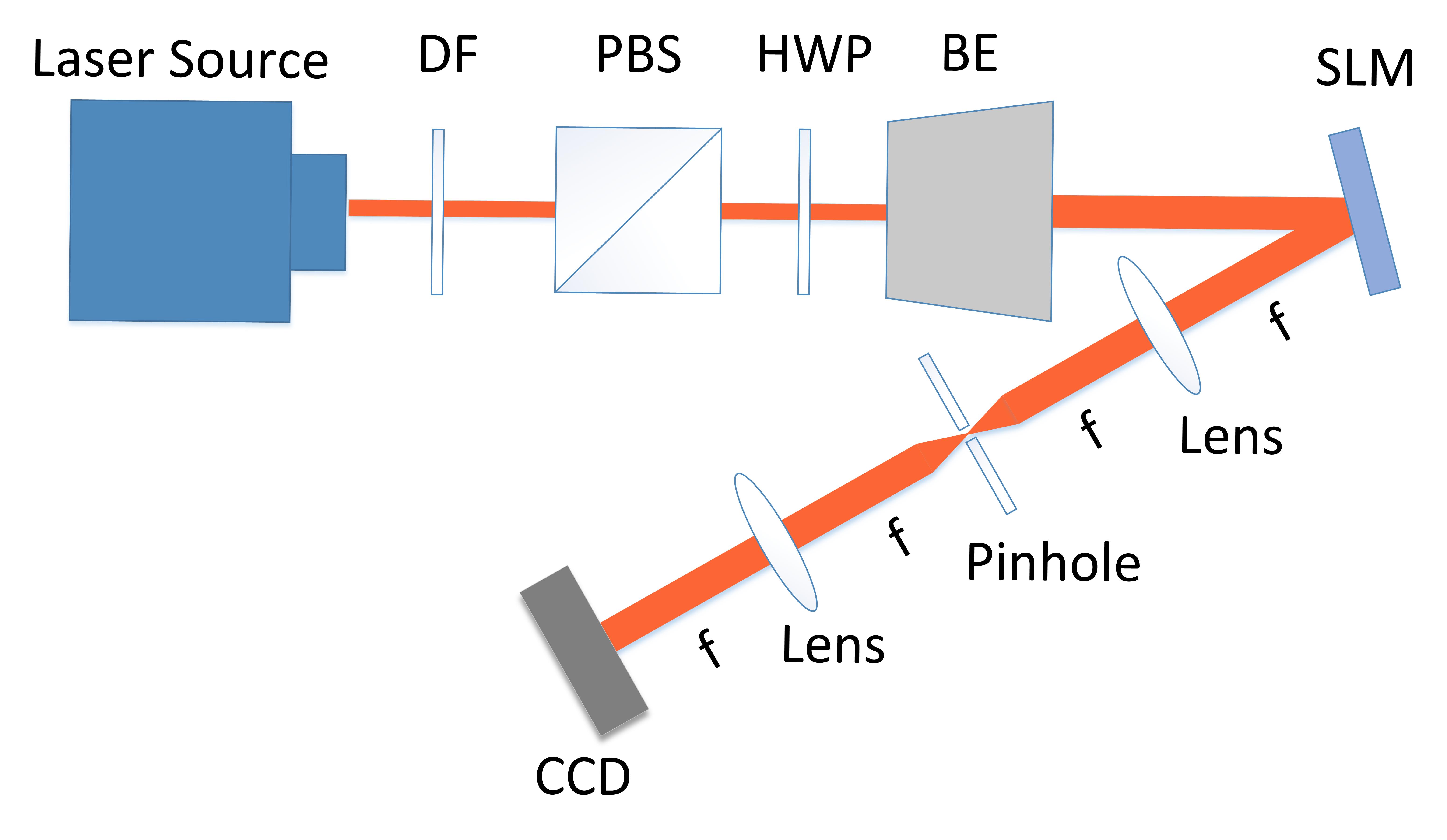}
	\caption{ The schematic of our experimental setup. DF, density filters; PBS, polarizing beam splitter; HWP, half wave plate; BE, beam expander; SLM, spatial light modulator; CCD, Charge-coupled device. The two lenses have the same focal length and f = 250 mm.}
	\label{fig-3}
\end{figure}

\begin{figure}[htbp]	
	\centering\includegraphics[width=\linewidth]{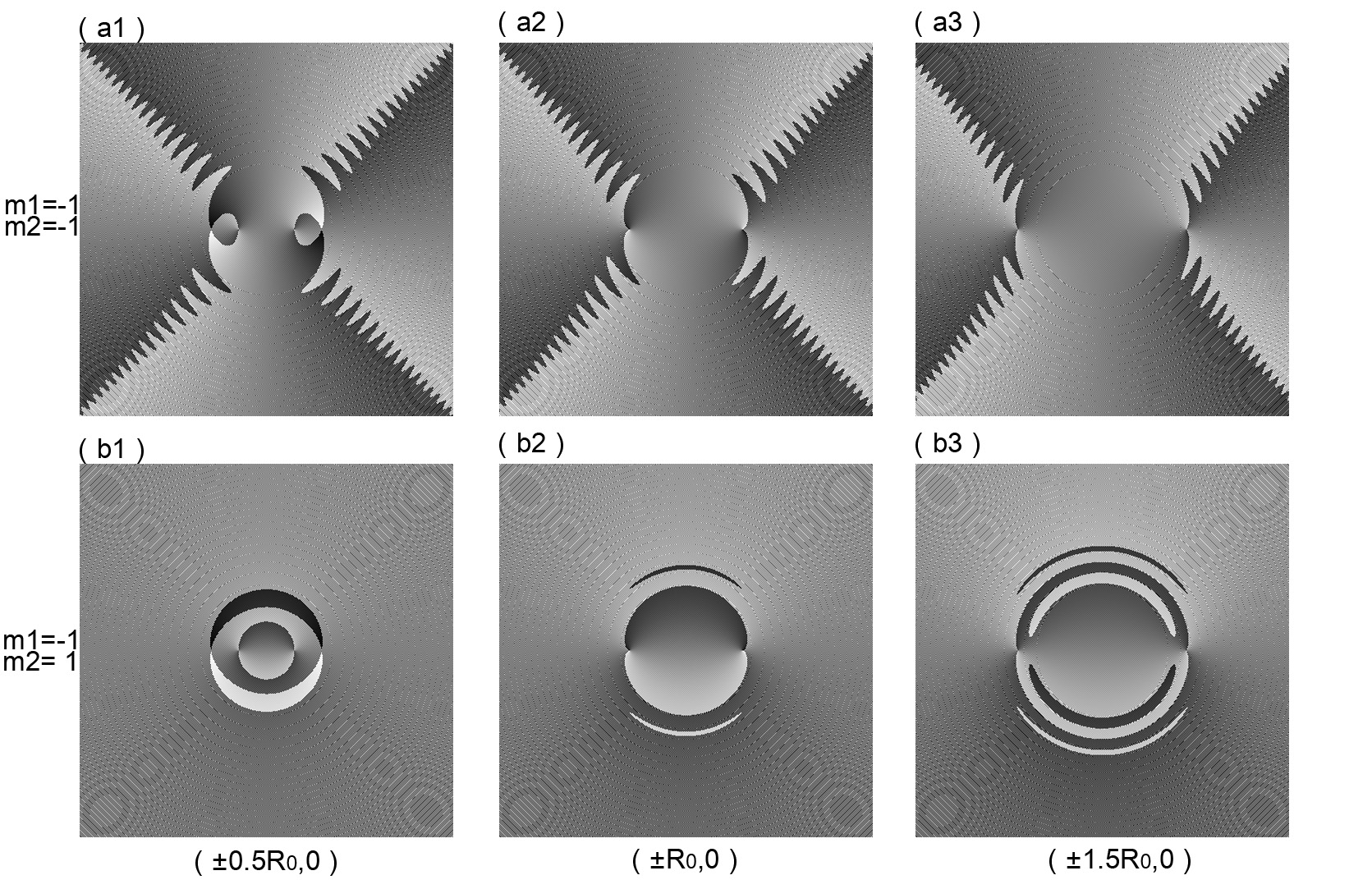}
	\caption{The designed phase masks without prism phase for CAVBs with a pair of vortices. (a) both the vortices have the same topological charge -1, $m_1=m_2=-1$; (b) the topological charges of the two vortices have different signs, with $m_1=-1,m_2=1$. The loading position is labelled below.}
	\label{fig-4}
\end{figure}

According to the simulating results and for simplicity, we only perform the experiments with $d_v=0.5R_0,1.0R_0,1.5R_0$. We use the checker-board method\cite{Mendoza-Yero:14} to generate the phase mask, load it onto the phase-only SLM (BNS, P512-1064-PCIe), and generate our beam through the 4f system. The schematic of our experimental setup is shown in Fig. \ref{fig-3}. The 1064 nm diode laser beam is attenuated by the DF, polarized by the PBS, and its polarization direction is adjusted by rotating the HWP to meet the SLM's requirements for the polarization direction of the incident light. After the beam expander, a quasi-plane wave is achieved. The designed phase mask with proper prism phase is loaded onto the SLM and the desired part of the reflected beam is passed through the pinhole at the Fourier plane of the first lens and the target beam is generated at the second lens' back focal plane (initial plane of the generated beam). The designed phase masks without prism phase are shown in Fig. \ref{fig-4}.

\begin{figure}[tb]	
	\centering\includegraphics[width=\linewidth]{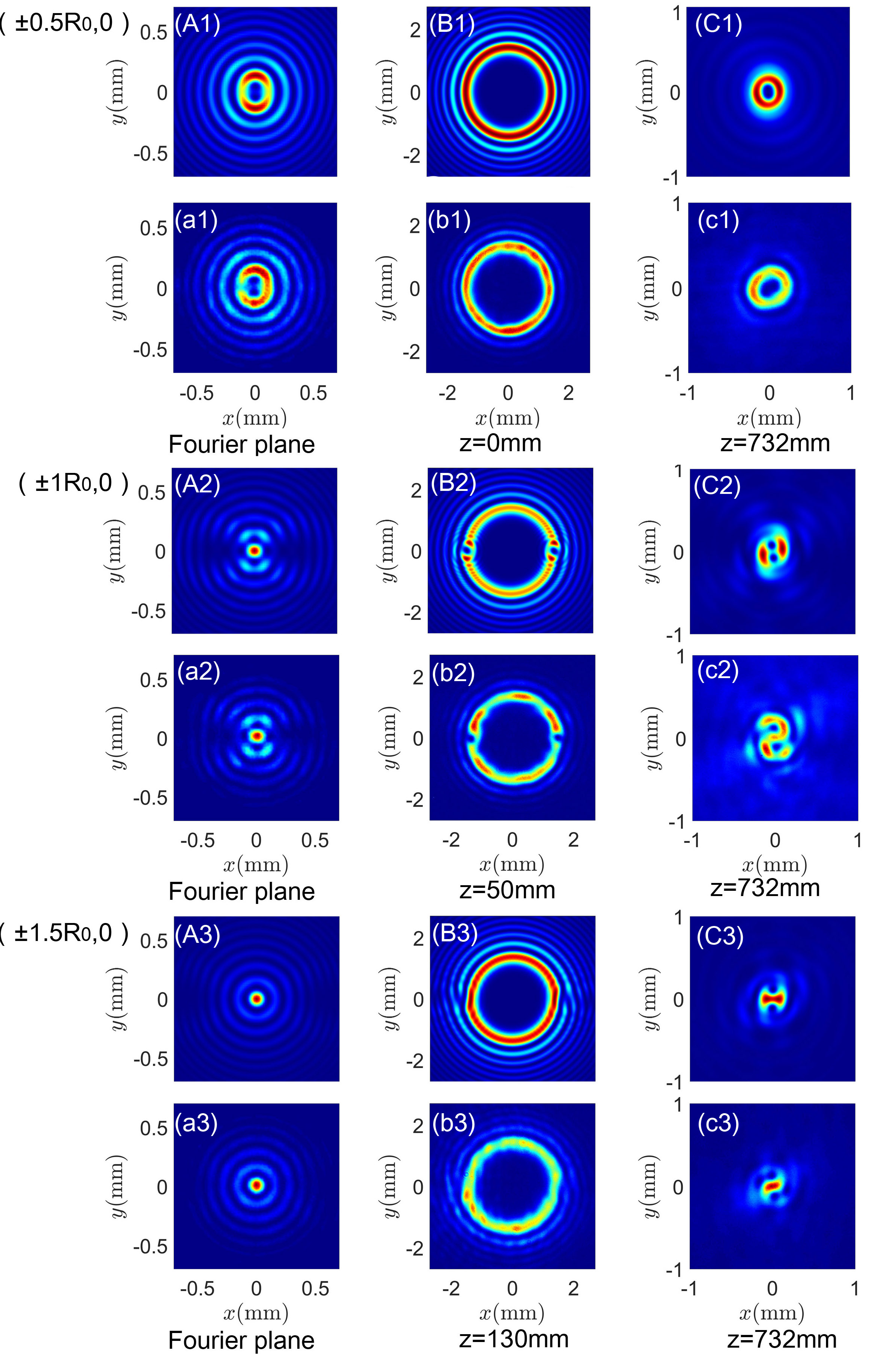}
	\caption{A comparison of simulating and experimental results of CAVBs carrying symmetrically a pair of vortices with same charges on the $x$-axis of the initial plane. All images basically have a point reflection symmetry through the origin.}
	\label{fig-5}
\end{figure}

\begin{figure}[hbp]	
	\centering\includegraphics[width=\linewidth]{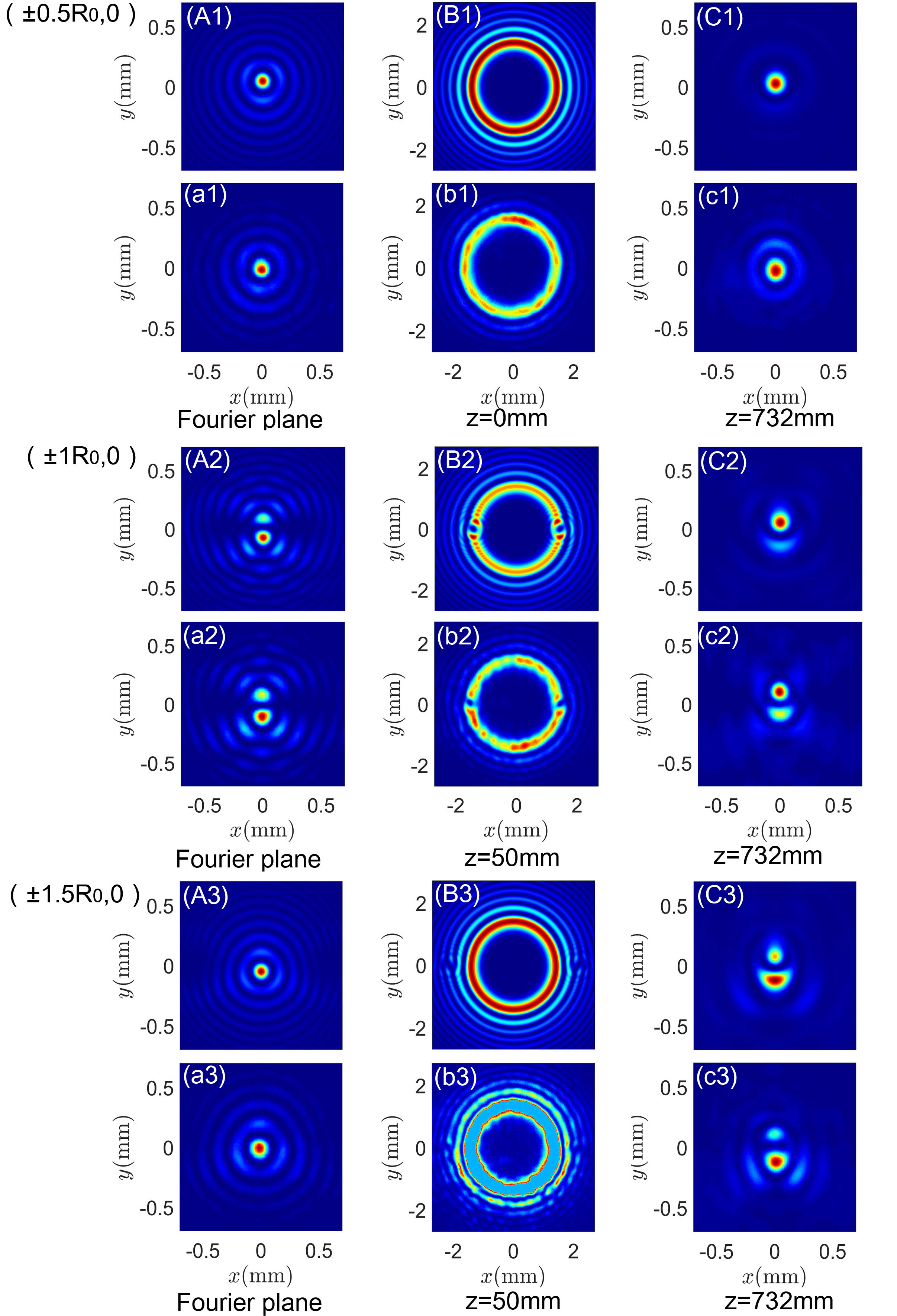}
	\caption{A comparison of simulating and experimental results of CAVBs carrying symmetrically a pair of vortices with opposite charges on the $x$-axis of the initial plane. All images are axisymmetric about the $y$-axis.}
	\label{fig-6}
\end{figure}

The experimental results are shown in Fig.\ref{fig-5} and Fig.\ref{fig-6}. The $d_v$ is used as the grouping basis and the position of two vortices is labelled at top-left of each group. The first line in each group are the simulating results, numbered with capital letters, and the second line are the experimental results, numbered with lowercase letters. The position marked below the second line in each group represents the position in the light path. The "Fourier plane" means the back focal plane of the first lens. The "z = 0" means the back focal plane of the second lens. The positions  corresponding to the second column in each group are different, and the position is selected for the purpose of showing the consistency of the experimental and simulating results as much as possible. The position of third column corresponds to the autofocusing plane of the CAVBs.

As shown in Fig.\ref{fig-5}, for the case of CAVBs carrying vortices with the same charges on the x-axis symmetrically in the initial plane, all images basically have a point reflection symmetry through the origin. The first column in Fig.\ref{fig-5}, which has not been mentioned in the theoretical analysis, represents the intensity distribution of the Fourier spectrum of the beam. The spectrum of the target beam has been achieved on the back focal plane of the first lens\cite{Mendoza-Yero:14}. For the sake of the spectrum of the CAVBs can be separated using a small aperture from the unmodulated light arising from direct reflection, a proper linear phase should be added to the phasemask. Though (A1) and (a1) have some inconsistency in energy distribution, the spectral intensity patterns recorded in the three groups of experiments are highly consistent with the numerical results. The inconsistency may be caused by a slight deviation between the centre of the incident beam and the centre of the SLM or there is a modulation error in SLM itself. 

For the {\bf first} group of Fig.\ref{fig-5}, that is $d_v=0.5R_0$, we found that the light intensity patterns observed near the z = 0 plane are almost the same, so the $z=0$ plane is selected as the observing plane. It can be seen from Fig.\ref{fig-5} (b1) and (B1) that the results of experiment and simulation consistent well. Good consistency for the autofocusing intensity patterns are achieved as shown in Fig.\ref{fig-5} (c1) and (C1), and it should be noticed that the connection of the brightest end of the ring in the simulation diagram (C1) is inclined upward to the right which agrees with the direction of the ring in experimental diagram (c1), though there is a little difference in the hollow shape. The {\bf second} group of Fig.\ref{fig-5} is corresponding to $d_v=1R_0$, in other words, a pair of point vortices on the initial plane are symmetrically located on the main ring of CAB. Although it is said that the point vortex does not change the light intensity distribution at initial plane, it is difficult to observe no-notch pattern like  Fig.\ref{fig-5} (b1) in experiments, so we choose the observation plane at z=50mm for the case of $d_v$ = 1$R_0$. Comparing  Fig.\ref{fig-5} (b2) and (B2), from the perspective of the notch orientation (related to the sign of the topological charge, where the left side is up and the right side is down) and the symmetry of the image, the experiment and simulation are highly consistent. The experimental and simulating images of the autofocusing plane are basically consistent in view of the direction of the line connecting the two hollow points. The {\bf third} group of Fig.\ref{fig-5} denotes the case of $d_v=1.5R_0$, in other words, the two vortices are located out of the main ring of CAB. For similar reason we choose the observation plane at z=130mm. It can be found that Fig.\ref{fig-5} (b3) and (B3), as well as (c3) and (C3) have a good agreement.  

Fig.\ref{fig-6} corresponds to vortices with different charges, and all images are basically axisymmetric about the $y$-axis. When $d_v=0.5R_0$, it can be seen from first group of Fig.\ref{fig-6} that our experiment results have a good agreement with simulation and the autofocusing pattern (Fig.\ref{fig-6} (c1) and (C1)) is a solid circular (Bessel-shaped) spot. When comparing the Fig.\ref{fig-6} (b2) and (B2), from the perspective of the notch orientation (related to the sign of the topological charge, here both the left and the right side are up) and the symmetry of the image, the experiment and simulation are highly consistent. It can be seen from Fig.\ref{fig-6} (c2) and (C2) that the solid circular autofocusing pattern has changed into two spots with most of the energy contained in the above spot and the centre becomes dark. As shown in Fig.\ref{fig-6} (b3), when $d_v=1.5R_0$, in order to see clearly the vortices outside the main ring near the initial plane (approximately at the intersection of the third ring and the $x$-axis), the experimental picture was overexposed. The experimental image (b3) agrees well with the simulating image (B3) in view of the position of the two vortices. The image (c3) is also in good agreement with simulating (C3), and it can be seen that the spot below now contains most of the energy.

\section{Summary and discussion about potential applications}
In summary, we have numerically and experientially study the relation between the loading position $d_v$ of the two vortices and the behaviour of the intensity pattern and vortices distribution of the CAVB at its autofocusing plane. 

For the point vortices loaded at initial plane having the same charges, the light intensity pattern (actually the filed of the beam) of the CAVB has a point symmetry about the origin. When $d_v$ is small (E.g. 0.5$R_0$), the initial vortices tend to form "one-vortex" on the focal plane of 2-times charge and the focal spot has a central hollow. As $d_v$ increases, the two vortices at the focal plane gradually deviate from each other along the $y$-axis with the emergence of two vortices having opposite charges from them,  and the central hollow becomes 2 dark spots in the light intensity pattern. At last, when the initial two vortices are loaded far enough (E.g. $d_v$ = 4$R_0$) from the beam centre, the four vortices disappear and focal spot returns to a Bessel-shaped  spot. Where the new vortices come from and the disappearance of the four vortices are worthy of further research.
For the point vortices having the opposite sign, the beam is symmetrical about the $y$-axis. When $d_v$ is small, the two vortices do not appear at the centre of autofocusing plane and focal pattern is a nearly circular solid (Bessel-shaped) bright spot at the beam centre, but not strictly in the centre. When $d_v$ increases, the circular solid bright spot gradually moves along positive $y$-axis away from origin, with some energy transferring to a newborn spot below the $x$-axis and a dark gap appear in the beam centre. As $d_v$ keeps increasing, the spot below $x$-axis keeps accumulating more energy and move closer to the origin along $y$-axis. Finally when the two vortices at initial plane of the CAVB are loaded far enough, the focal spot returns to a Bessel-shaped spot. In practice, it may be possible to use the controlling effect of $d_v$ on the autofocusing behaviour to encode information for optical communication. In addition, the subtle adjustment effect that $d_v$ have on the bright or hollow spot position at the autofocusing plane may be helpful for particle capture researches and applications.

\section*{Funding}
National Key Research and Development Program of China (2017YFB0503100); National Natural Science Foundation of China (NSFC) (11474254, 11804298); Fundamental Research Funds for the Central Universities (2017QN81005, 2016XZZX004-01).

\section*{Disclosures}
The authors declare no conflicts of interest.


\balance
\bibliography{my}

\begin{thebibliography}{10}
\newcommand{\enquote}[1]{``#1''}

\bibitem{doi:10.1119/1.11855}
M.~V. Berry and N.~L. Balazs, \enquote{Nonspreading wave packets,}
  {\protect\JournalTitle{American Journal of Physics}} \textbf{47}, 264--267
  (1979).

\bibitem{Siviloglou:07}
G.~A. Siviloglou and D.~N. Christodoulides, \enquote{Accelerating finite energy
  airy beams,} {\protect\JournalTitle{Opt. Lett.}} \textbf{32}, 979--981
  (2007).

\bibitem{recent_advances_Efremidis:19}
N.~K. Efremidis, Z.~Chen, M.~Segev, and D.~N. Christodoulides, \enquote{Airy
  beams and accelerating waves: an overview of recent advances,}
  {\protect\JournalTitle{Optica}} \textbf{6}, 686--701 (2019).

\bibitem{Zhuang:15}
F.~Zhuang, Z.~Zhu, J.~Margiewicz, and Z.~Shi, \enquote{Quantitative study on
  propagation and healing of airy beams under experimental conditions,}
  {\protect\JournalTitle{Opt. Lett.}} \textbf{40}, 780--783 (2015).

\bibitem{Chu.2012}
X.~Chu, G.~Zhou, and R.~Chen, \enquote{Analytical study of the self-healing
  property of airy beams,} {\protect\JournalTitle{Physical Review A}}
  \textbf{85} (2012).

\bibitem{Efremidis:10}
N.~K. Efremidis and D.~N. Christodoulides, \enquote{Abruptly autofocusing
  waves,} {\protect\JournalTitle{Opt. Lett.}} \textbf{35}, 4045--4047 (2010).

\bibitem{Pre-engineered-Chremmos:11}
I.~Chremmos, N.~K. Efremidis, and D.~N. Christodoulides,
  \enquote{Pre-engineered abruptly autofocusing beams,}
  {\protect\JournalTitle{Opt. Lett.}} \textbf{36}, 1890--1892 (2011).

\bibitem{Papazoglou:11}
D.~G. Papazoglou, N.~K. Efremidis, D.~N. Christodoulides, and S.~Tzortzakis,
  \enquote{Observation of abruptly autofocusing waves,}
  {\protect\JournalTitle{Opt. Lett.}} \textbf{36}, 1842--1844 (2011).

\bibitem{Chremmos:11}
I.~Chremmos, P.~Zhang, J.~Prakash, N.~K. Efremidis, D.~N. Christodoulides, and
  Z.~Chen, \enquote{Fourier-space generation of abruptly autofocusing beams and
  optical bottle beams,} {\protect\JournalTitle{Opt. Lett.}} \textbf{36},
  3675--3677 (2011).

\bibitem{Zhang:s}
P.~Zhang, J.~Prakash, Z.~Zhang, M.~S. Mills, N.~K. Efremidis, D.~N.
  Christodoulides, and Z.~Chen, \enquote{Trapping and guiding microparticles
  with morphing autofocusing airy beams,} {\protect\JournalTitle{Opt. Lett.}}
  \textbf{36}, 2883--2885 (2011).

\bibitem{Jiang:s1}
Y.~Jiang, K.~Huang, and X.~Lu, \enquote{Radiation force of abruptly
  autofocusing airy beams on a rayleigh particle,} {\protect\JournalTitle{Opt.
  Express}} \textbf{21}, 24413--24421 (2013).

\bibitem{Jiang:s2}
Y.~Jiang, Z.~Cao, H.~Shao, W.~Zheng, B.~Zeng, and X.~Lu, \enquote{Trapping two
  types of particles by modified circular airy beams,}
  {\protect\JournalTitle{Opt. Express}} \textbf{24}, 18072--18081 (2016).

\bibitem{Wang2014}
F.~Wang, C.~Zhao, Y.~Dong, Y.~Dong, and Y.~Cai, \enquote{Generation and
  tight-focusing properties of cylindrical vector circular airy beams,}
  {\protect\JournalTitle{Applied Physics B}} \textbf{117}, 905--913 (2014).

\bibitem{Panagiotopoulos2013}
P.~Panagiotopoulos, D.~G. Papazoglou, A.~Couairon, and S.~Tzortzakis,
  \enquote{Sharply autofocused ring-airy beams transforming into non-linear
  intense light bullets,} {\protect\JournalTitle{Nature Communications}}
  \textbf{4}, 2622 (2013).

\bibitem{Manousidaki:s}
M.~Manousidaki, D.~G. Papazoglou, M.~Farsari, and S.~Tzortzakis,
  \enquote{Abruptly autofocusing beams enable advanced multiscale
  photo-polymerization,} {\protect\JournalTitle{Optica}} \textbf{3}, 525--530
  (2016).

\bibitem{COULLET1989403}
P.~Coullet, L.~Gil, and F.~Rocca, \enquote{Optical vortices,}
  {\protect\JournalTitle{Optics Communications}} \textbf{73}, 403 -- 408
  (1989).

\bibitem{He.1995}
He, Friese, Heckenberg, and Rubinsztein-Dunlop, \enquote{Direct observation of
  transfer of angular momentum to absorptive particles from a laser beam with a
  phase singularity,} {\protect\JournalTitle{Physical review letters}}
  \textbf{75}, 826--829 (1995).

\bibitem{Gahagan.1996}
K.~T. Gahagan and G.~A. Swartzlander, \enquote{Optical vortex trapping of
  particles,} {\protect\JournalTitle{Optics Letters}} \textbf{21}, 827--829
  (1996).

\bibitem{ZhongxiWang.2011}
{Zhongxi Wang}, {N. Zhang}, and {X.-C. Yuan}, \enquote{High-volume optical
  vortex multiplexing and de-multiplexing for free-space optical
  communication,} {\protect\JournalTitle{Optics Express}} \textbf{19}, 482--492
  (2011).

\bibitem{Gibson.2004}
G.~Gibson, J.~Courtial, M.~Padgett, M.~Vasnetsov, V.~Pas'ko, S.~Barnett, and
  S.~Franke-Arnold, \enquote{Free-space information transfer using light beams
  carrying orbital angular momentum,} {\protect\JournalTitle{Optics express}}
  \textbf{12}, 5448--5456 (2004).

\bibitem{Lavery.2017}
M.~P.~J. Lavery, C.~Peuntinger, K.~G{\"u}nthner, P.~Banzer, D.~Elser, R.~W.
  Boyd, M.~J. Padgett, C.~Marquardt, and G.~Leuchs, \enquote{Free-space
  propagation of high-dimensional structured optical fields in an urban
  environment,} {\protect\JournalTitle{Science advances}} \textbf{3}, e1700552
  (2017).

\bibitem{G.Anzolin.2009}
{G. Anzolin}, {F. Tamburini}, {A. Bianchini}, and {C. Barbieri},
  \enquote{Method to measure off-axis displacements based on the analysis of
  the intensity distribution of a vortex beam,} {\protect\JournalTitle{Physical
  Review A}} \textbf{79}, 033845 (2009).

\bibitem{MichaelMazilu.2009}
M.~Mazilu, J.~Baumgartl, T.~Čižmár, and K.~Dholakia, \enquote{{Accelerating
  vortices in Airy beams},} in \emph{Laser Beam Shaping X,}  vol. 7430
  A.~Forbes and T.~E. Lizotte, eds., International Society for Optics and
  Photonics (SPIE, 2009), pp. 68 -- 75.

\bibitem{Dai:10}
H.~T. Dai, Y.~J. Liu, D.~Luo, and X.~W. Sun, \enquote{Propagation dynamics of
  an optical vortex imposed on an airy beam,} {\protect\JournalTitle{Opt.
  Lett.}} \textbf{35}, 4075--4077 (2010).

\bibitem{Dai:11}
H.~T. Dai, Y.~J. Liu, D.~Luo, and X.~W. Sun, \enquote{Propagation properties of
  an optical vortex carried by an airy beam: experimental implementation,}
  {\protect\JournalTitle{Opt. Lett.}} \textbf{36}, 1617--1619 (2011).

\bibitem{Chen.2011}
R.-P. Chen and C.~H.~R. Ooi, \enquote{Nonclassicality of vortex airy beams in
  the wigner representation,} {\protect\JournalTitle{Physical Review A}}
  \textbf{84} (2011).

\bibitem{Chen.2012}
R.-P. Chen, L.-X. Zhong, Q.~Wu, and K.-H. Chew, \enquote{Propagation properties
  and m2 factors of a vortex airy beam,} {\protect\JournalTitle{Optics {\&}
  Laser Technology}} \textbf{44}, 2015--2019 (2012).

\bibitem{Davis:12}
J.~A. Davis, D.~M. Cottrell, and D.~Sand, \enquote{Abruptly autofocusing vortex
  beams,} {\protect\JournalTitle{Opt. Express}} \textbf{20}, 13302--13310
  (2012).

\bibitem{Davis:13-3}
J.~A. Davis, D.~M. Cottrell, and J.~M. Zinn, \enquote{Direct generation of
  abruptly focusing vortex beams using a 3/2 radial phase-only pattern,}
  {\protect\JournalTitle{Appl. Opt.}} \textbf{52}, 1888--1891 (2013).

\bibitem{Jiang:12}
Y.~Jiang, K.~Huang, and X.~Lu, \enquote{Propagation dynamics of abruptly
  autofocusing airy beams with optical vortices,} {\protect\JournalTitle{Opt.
  Express}} \textbf{20}, 18579--18584 (2012).

\bibitem{Dongmei-Chen:s}
B.~Chen, C.~Chen, X.~Peng, Y.~Peng, M.~Zhou, and D.~Deng, \enquote{Propagation
  of sharply autofocused ring airy gaussian vortex beams,}
  {\protect\JournalTitle{Opt. Express}} \textbf{23}, 19288--19298 (2015).

\bibitem{Jiang:s33}
Y.~Jiang, S.~Zhao, W.~Yu, and X.~Zhu, \enquote{Abruptly autofocusing property
  of circular airy vortex beams with different initial launch angles,}
  {\protect\JournalTitle{J. Opt. Soc. Am. A}} \textbf{35}, 890--894 (2018).

\bibitem{JingliZhuang.2020}
{Jingli Zhuang}, {Liping Zhang}, and {Dongmei Deng}, \enquote{Tight-focusing
  properties of linearly polarized circular airy gaussian vortex beam,}
  {\protect\JournalTitle{Optics Letters}} \textbf{45}, 296--299 (2020).

\bibitem{XiangZhang.2020}
{Xiang Zhang}, {Peng Li}, {Sheng Liu}, {Bingyan Wei}, {Shuxia Qi}, {Xinhao
  Fan}, {Shouheng Wang}, {Yuan Zhang}, and {Jianlin Zhao},
  \enquote{Autofocusing of ring airy beams embedded with off-axial vortex
  singularities,} {\protect\JournalTitle{Optics Express}} \textbf{28},
  7953--7960 (2020).

\bibitem{Mendoza-Yero:14}
O.~Mendoza-Yero, G.~M\'{i}nguez-Vega, and J.~Lancis, \enquote{Encoding complex
  fields by using a phase-only optical element,} {\protect\JournalTitle{Opt.
  Lett.}} \textbf{39}, 1740--1743 (2014).

\bibitem{D.R.1997}
{D. Roza, C. T. Law, G. A. Swartzlander, Jr.}, \enquote{Propagation dynamics of
  optical vortices,} {\protect\JournalTitle{J. Opt. Soc. Am. B, JOSAB (JOSA
  B)}} \textbf{14}, 3054--3065 (1997).

\bibitem{Li:14}
N.~Li, Y.~Jiang, K.~Huang, and X.~Lu, \enquote{Abruptly autofocusing property
  of blocked circular airy beams,} {\protect\JournalTitle{Opt. Express}}
  \textbf{22}, 22847--22853 (2014).

\bibitem{XIE2018288}
W.~Xie, P.~Zhang, H.~Wang, and X.~Chu, \enquote{Propagation of a vortex
  elliptical airy beam,} {\protect\JournalTitle{Optics Communications}}
  \textbf{427}, 288 -- 293 (2018).

\end{thebibliography}

\bibliographyfullrefs{my}


\end{document}